\documentclass[cameraready]{Interspeech}

\title{What and When to Learn: CURriculum Ranking Loss for Large-Scale Speaker Verification}

\author[]{Massa}{Baali}
\author[]{Sarthak}{Bisht}
\author[]{Rita}{Singh}
\author[]{Bhiksha}{Raj}

\address{
    Carnegie Mellon University, Pittsburgh, USA
}

\email{mbaali@cs.cmu.edu}

\usepackage{comment}
\usepackage{amsmath}
\usepackage{multirow}
\usepackage{graphicx}
\usepackage{subcaption}
\usepackage[ruled,vlined]{algorithm2e}

\begin{document}

\maketitle

\begin{abstract}
    Speaker verification at large scale remains an open challenge as fixed-margin losses treat all samples equally regardless of quality. We hypothesize that mislabeled or degraded samples introduce noisy gradients that disrupt compact speaker manifolds. We propose Curry (CURriculum Ranking), an adaptive loss that estimates sample difficulty online via Sub-center ArcFace: confidence scores from dominant sub-center cosine similarity rank samples into easy, medium, and hard tiers using running batch statistics, without auxiliary annotations. Learnable weights guide the model from stable identity foundations through manifold refinement to boundary sharpening. To our knowledge, this is the largest-scale speaker verification system trained to date. Evaluated on VoxCeleb1-O, and SITW, Curry reduces EER by 86.8\% and 60.0\% over the Sub-center ArcFace baseline, establishing a new paradigm for robust speaker verification on imperfect large-scale data.
\end{abstract}

\section{Introduction}
Recently, most speaker verification models \cite{snyder2018x, vera2015speaker,baali2025delulu, doddipatla2017speaker,baali2024pdaf} have transitioned into the era of large-scale recognition, with systems now scaling to hundreds of thousands of identities \cite{lin2024voxblink2, yakovlev2024reshape}. While this expansion is essential for real-world robustness, it introduces a fundamental conflict between data volume and training stability. In regimes with high acoustic variability and diverse data sources, the acoustic diversity of the training data creates a significant \textit{gradient noise}, where the model is forced to learn clean, representative utterances alongside highly degraded or mislabeled samples. This struggle to filter noise amidst vast data volumes hinders the model's ability to converge on robust representations \cite{jung2022large}.
Conventional loss functions, such as standard AAM-Softmax \cite{deng2019arcface}, impose uniform margins and gradient updates across all samples, treating a high-confidence, clean recording identically to a noisy, ambiguous one. We observe that this uniform treatment undermines the formation of compact speaker manifolds; the model implicitly learns from \textit{hard} or corrupted samples too early, which disrupts the decision boundaries of nearby identities.

Futhermore, curriculum learning~\cite{bengio2009curriculum} offers a principled mechanism to
circumvent this problem by controlling the order and timing of training samples.
However, existing curriculum strategies often focus exclusively on defining what data
to learn, while neglecting when that data should be introduced (the pacing). Moreover, this definition of \textbf{what} is often static; they treat difficulty as an inherent property of the data rather than a dynamic state that should adapt to the model’s evolving maturity. Consequently, existing methods fail to synchronize data complexity with the model’s convergence, causing them to be ineffective for the dynamic requirements of training on large-scale identities. 

To overcome these barriers, we introduce an adaptive curriculum learning framework that simultaneously determines what to learn and when to introduce complexity, while remaining fully responsive to the model's evolving state. Our framework identifies sample complexity on the fly, without requiring auxiliary difficulty labels. By leveraging the geometry of Sub-centerArcFace~\cite{deng2020sub}  which maintains multiple prototype vectors per speaker class to capture intra-class acoustic variability; we derive per-sample confidence scores from the dominant sub-center cosine similarity. These scores dynamically partition each mini-batch into tiered levels of difficulty that adjust as the model matures. We propose a structured training trajectory where the model regulates its own learning pace: it first establishes robust identity foundations using clean samples, systematically introduces variations, and finally focuses on confusable boundaries. Learnable weights modulate gradient contributions, allowing the model to adapt its learning schedule to its internal progress.

Our primary contributions are as follows:
\begin{itemize}
    \item \textbf{Curry Loss:} We introduce \textit{Curry} (Curriculum Ranking), a
    novel loss function that wraps any differentiable per-sample objective with
    adaptive, tier-based gradient weighting. Curry is loss-agnostic by design; it
    can be combined with AAM-Softmax, Sub-center ArcFace, or any future speaker loss
    without architectural changes, making it a general-purpose curriculum wrapper for
    large-scale training.

    \item \textbf{Difficulty Ranking:} We propose an unsupervised scoring
    mechanism based on sub-center angular distances that identifies sample corruption
    risk dynamically and partitions training data into difficulty tiers using running
    batch statistics, eliminating the need for manual or offline difficulty
    annotations.

    \item \textbf{Large-Scale System:} To our knowledge, we present the
    largest-scale speaker verification system trained to date, spanning 500K+
    identities across VoxCeleb, VoxBlink2, and CommonVoice. Evaluated on SVeritas,
    VoxCeleb1-O, and SITW, Curry reduces EER by 86.8\% and 60.0\% over the
    Sub-center ArcFace baseline respectively.
\end{itemize}


\section{Literature Review}
Scaling speaker verification to hundreds of thousands of identities introduces fundamental tensions between data volume and training stability. Jung et al. \cite{jung2022large} showed that not all architectures benefit equally from data scaling when combining VoxCeleb, NIST SRE, and CommonVoice (up to 87,000 speakers), and that managing data quality becomes critical at scale. Singh and Raj \cite{singh2025human} provided theoretical grounding for this challenge: while their analysis of 44 causally independent acoustic features confirms that large-scale speaker verification is fundamentally feasible, it also implies that preserving discriminative information under real-world degradation is the central bottleneck.

The severity of this bottleneck grows with data scale. Ahmed and Imtiaz \cite{ahmed2026quantifying} showed that quality metrics such as PESQ explain up to 69\% of EER variance for SSL-based models, leaving a substantial portion attributable to factors beyond simple acoustic degradation. The problem is compounded by label noise: the semi-automated cleaning pipeline of CommonBench \cite{hintz2024commonbench} revealed significant label corruption in crowdsourced CommonVoice data, and Farhadipour et al. \cite{farhadipour2026tidyvoice} found that approximately 9\% of multilingual speakers had identity switches across languages. Together, these findings establish that large-scale training data is inherently noisy along both acoustic and label dimensions, motivating training strategies that can adapt to sample reliability. A complementary direction addresses this scarcity at the speaker level rather than the sample level: Baali et al.\ \cite{baali2025caarma} proposed CAARMA, which augments the training set with synthetic speaker identities via adversarial mixup in the embedding space, effectively expanding the number of available classes rather than individual samples. While orthogonal to our approach, this highlights a growing recognition that both sample quality and class diversity are critical bottlenecks at scale.

Curriculum learning \cite{bengio2009curriculum} offers a principled framework for such adaptation by structuring training from easy to hard examples. Two dimensions define any curriculum strategy: (1) \textbf{what to learn}: how sample difficulty is defined and which samples are selected, and (2) \textbf{when to learn}: the pacing function that determines when harder examples are introduced. Several works have applied curriculum learning to speaker recognition: Ranjan and Hansen \cite{ranjan2017curriculum} developed curriculum-based algorithms for noise-robust speaker recognition at the i-vector and PLDA stages; Heo et al. \cite{heo2022curriculum} proposed dataset-level and augmentation-level curricula for self-supervised verification within the DINO framework; and Bai et al. \cite{bai2022end} introduced a curriculum bipartite ranking approach at the loss function level.

However, these approaches share key limitations. They typically define difficulty statically (based on data properties or pre-computed scores) rather than dynamically adapting to the model's evolving internal state. Furthermore, none simultaneously addresses both the \textbf{`what'} and \textbf{`when'} dimensions in a unified framework, nor has any been demonstrated at the scale of hundreds of thousands of identities where noise management is most critical.
\section{Adaptive Curriculum Framework}
\begin{figure}[t]
  \centering
  \includegraphics[width=1.00\columnwidth, trim=10 5 10 5, clip]{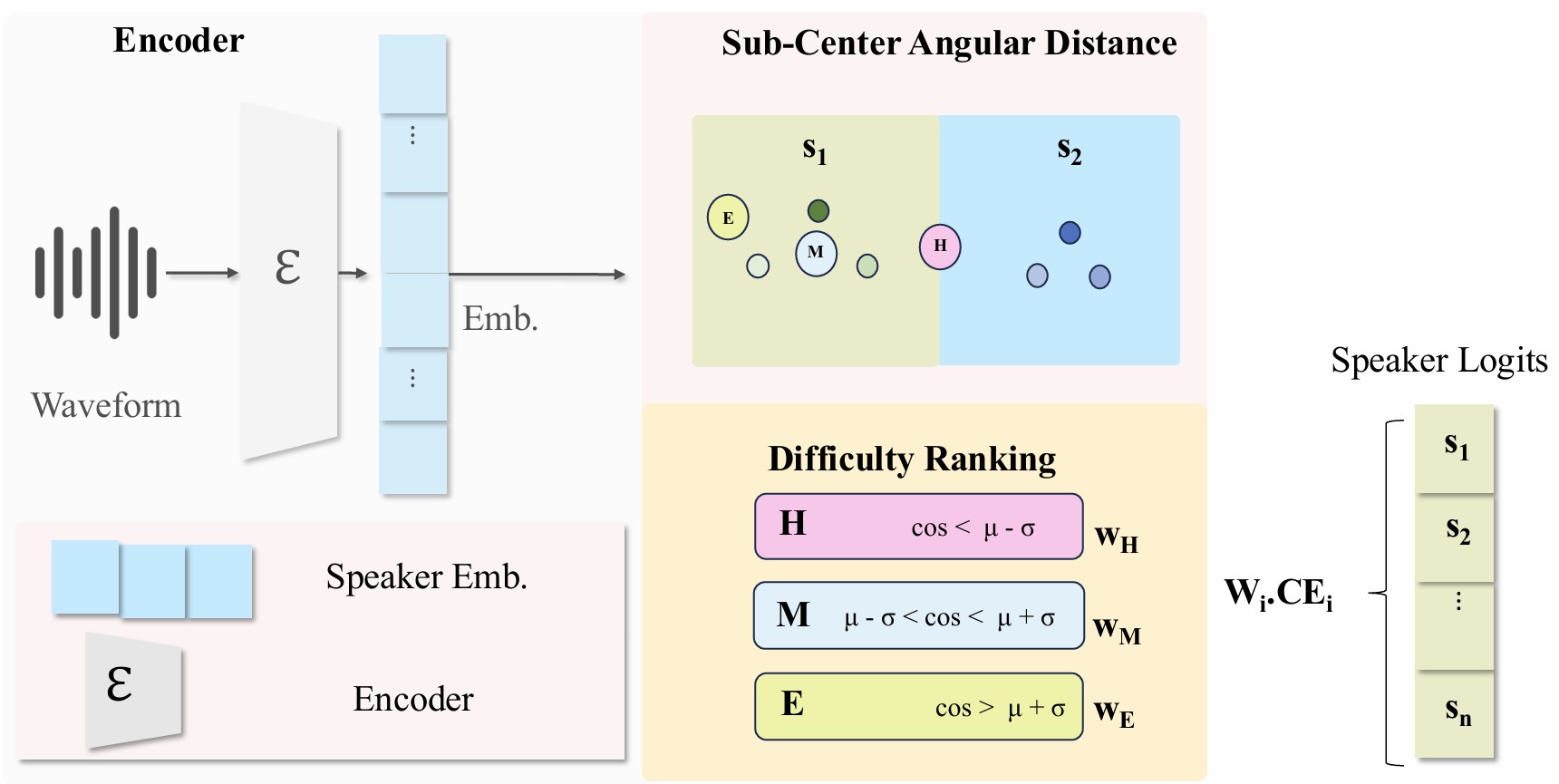}
  \caption{The adaptive curriculum learning pipeline. The encoder
  $\mathcal{E}$ maps raw waveforms to speaker embeddings projected into a
  sub-center angular distance space, where each speaker region contains $K$
  sub-centers capturing acoustic variability. Per-sample confidence scores
  from the dominant sub-center cosine similarity are ranked against running
  batch statistics $(\hat{\mu}, \hat{\sigma})$ into Hard, Medium, and Easy
  tiers. Learnable weights $W_H$, $W_M$, $W_E$ scale each sample's
  gradient contribution before loss reduction.}
  \label{fig:pipeline}
\end{figure}

This section describes our proposed framework as shown in Figure \ref{fig:pipeline}. We first introduce the speaker encoder and its feature extraction mechanism, then formalize the sub-center angular distance scoring that drives difficulty estimation, and finally present the Curry loss that integrates all components.

\subsection{Speaker Encoder}
We adopt W2V-BERT~2.0~\cite{chung2021w2v} as our speaker encoder $\varepsilon$. W2V-BERT~2.0 is a large-scale self-supervised model trained on 4.5 million hours of unlabeled audio using a hybrid objective that combines masked prediction and contrastive learning. 

Given a raw waveform $\mathbf{x} \in \mathbb{R}^{T}$, we first extract its log-Mel spectrograms and pass them into the pre-trained W2V-BERT~2.0 to obtain the hidden representations
of each layer. The model processes the input through 24 Conformer layers, producing a
sequence of hidden states $\{\mathbf{h}_l\}_{l=0}^{L}$, where $L=24$.

To aggregate complementary speaker-discriminative information across all layers, we adopt
a layer-wise weighted average~\cite{chen2022wavlm} following~\cite{li2025enhancing}, where each layer is assigned a
learnable scalar weight updated during training. The final frame-level feature is obtained by computing a softmax-normalized weighted sum of all layer outputs, allowing
the network to selectively emphasize layers that carry the most speaker-relevant information.

The resulting frame sequence is then passed to the MFA backend, which applies
a lightweight adapter module to each layer output before aggregation, followed by Attentive
Statistics Pooling (ASP)~\cite{okabeattentive}. ASP computes an attention-weighted mean and standard
deviation over the temporal dimension, capturing both the average speaker characteristics
and their variability across frames. The pooled statistics are concatenated and projected
through a fully-connected layer with batch normalization, yielding a fixed-dimensional
speaker embedding $\mathbf{e} \in \mathbb{R}^{d}$, with $d=192$ in our experiments.

\subsection{Difficulty Ranking}

A central observation motivating our approach is that not all training samples carry
equally reliable gradient signal. In large-scale settings that incorporate diverse data
sources, many utterances are mislabeled, highly degraded, or acoustically ambiguous.
Treating such samples uniformly alongside clean recordings disrupts the formation of
compact speaker manifolds, causing the model to learn from corrupted or noisy samples
too early and undermining the stability of the optimization.

\subsubsection{Confidence Scoring via Sub-center Angular Distance}

To quantify per-sample reliability without any auxiliary annotations, we leverage the
geometry of Sub-center ArcFace~\cite{deng2020sub}. Instead of representing each
speaker class $y$ by a single prototype, we maintain $K$ sub-center weight vectors
$\{\mathbf{c}_y^k\}_{k=1}^{K}$, each capturing a distinct acoustic condition within the
class. The dominant sub-center for a given sample is the one with the highest cosine
similarity to its embedding, and the corresponding cosine value serves as a natural
per-sample confidence score. Formally, for an embedding $\mathbf{e}_i$, the
\textbf{target logit} is defined as:
\begin{equation}
s_i 
= \max_{k \in \{1,\ldots,K\}} \cos(\theta_{y_i}^k)
= \max_{k \in \{1,\ldots,K\}}
\frac{\mathbf{e}_i \cdot \mathbf{c}_{y_i}^k}
     {\lVert \mathbf{e}_i \rVert \,\lVert \mathbf{c}_{y_i}^k \rVert}.
\label{eq:target_logit}
\end{equation}
The intuition is straightforward: a sample well-aligned with its speaker's dominant
sub-center yields a high $s_i$, indicating a clean and unambiguous utterance, while a
degraded or mislabeled sample drifts toward a non-dominant sub-center, yielding a low
or negative $s_i$. Sub-centers therefore implicitly factorize acoustic conditions within
a class e.g., clean speech, reverberated, and noisy recordings each tend to cluster around
different sub-centers without requiring any explicit condition labels. We set $K=3$
in all experiments.

\subsubsection{Dynamic Tier Assignment via Running Batch Statistics}

Rather than computing a global difficulty ranking over the entire dataset, which would
be extremely expensive at our scale (500K speakers), we estimate difficulty locally
within each mini-batch using exponential moving averages of the target logit distribution:
\begin{equation}
    \hat{\mu} \leftarrow (1-m)\,\hat{\mu} + m \cdot \mu_b, \qquad
    \hat{\sigma} \leftarrow (1-m)\,\hat{\sigma} + m \cdot \sigma_b,
    \label{eq:running_stats}
\end{equation}
where $\mu_b$ and $\sigma_b$ are the mean and standard deviation of $\{s_i\}$ within
the current batch, and $m=0.01$ is the momentum coefficient. These statistics evolve
continuously as training progresses, automatically adapting the difficulty thresholds to
the model's improving representations; a key \textit{advantage} over static, offline difficulty
labels that cannot respond to the model's internal state.

Each sample is then assigned to one of three tiers based on its target logit relative
to the running statistics:
\begin{equation}
    \mathrm{tier}(i) = \begin{cases}
        \mathrm{Easy}   & \text{if } s_i > \hat{\mu} + \hat{\sigma}, \\
        \mathrm{Hard}   & \text{if } s_i < \hat{\mu} - \hat{\sigma}, \\
        \mathrm{Medium} & \text{otherwise.}
    \end{cases}
    \label{eq:tiers}
\end{equation}
This partitioning is computed at no additional forward-pass cost, as $s_i$ is a direct byproduct of the Sub-center ArcFace computation. 
\subsection{Curry Loss}
We formalize the adaptive curriculum weighting as a standalone loss function, which we name \textbf{Curry} (\textbf{Cur}riculum \textbf{r}anking with d\textbf{y}namic weighting). Curry is designed as a general-purpose wrapper: given any per-sample loss function $L$,
$\mathcal{L}$ (e.g., Sub-center ArcFace \cite{deng2020sub}), the Curry loss $\mathcal{L}_{Curry}$ is defined as the weighted aggregation of individual sample losses:

\begin{equation}
    \mathcal{L}_{Curry} = \frac{1}{|\mathcal{B}|} \sum_{i \in \mathcal{B}} w_i \cdot \mathcal{L}(\mathbf{e}_i, y_i)
\end{equation}

where $w_i \in \{W_{easy}, W_{medium}, W_{hard}\}$ are the tier-specific weights derived from the softmax-normalized curriculum logits $\boldsymbol{\gamma}$. By applying these weights, the \textit{Curry} loss prevents the encoder from being overwhelmed by noisy, ambiguous samples during the early stages of training, effectively decoupling the learning of robust identity foundations from the noise-handling characteristic of later training phases.

The evolution of these weights follows a three-phase schedule synchronized with the
encoder's maturity, as summarized in Algorithm~\ref{algo:curr}. In Phase~I, gradient
flow is restricted to Easy samples ($W_{\mathrm{medium}}, W_{\mathrm{hard}} \approx 0$)
to establish clean identity structure. Phase~II unlocks the Medium tier, allowing the encoder to map acoustic variations onto the stabilized centroids. In Phase~III, the Hard
tier is activated and $\boldsymbol{\gamma}$ becomes learnable, enabling the model to
refine decision boundaries between confusable speakers.

Unlike static curriculum approaches, \textit{Curry} continuously adapts to the model's internal state. As the encoder matures, $\hat{\mu}$ rises and $\hat{\sigma}$ tightens, automatically shifting the tier boundaries upward. Samples previously ranked as Hard naturally moves to Medium or Easy as the encoder learns more robust speaker features. This creates an adaptive feedback loop in which the model's own improving geometry determines which samples still challenge it, ensuring that gradient pressure is always matched to the model's current capacity to resolve speaker identities.

\section{Experimental Setup}

\begin{algorithm}[t]
\small
\caption{Curry Loss}
\label{algo:curr}

\KwIn{Encoder $E$, Loss $L$, Epoch $t$, Dataset $D$}
\KwOut{Updated parameters $\theta$}

Initialize $\boldsymbol{\mu}_s \leftarrow 0$, 
$\boldsymbol{\sigma}_s \leftarrow 1$, 
$\boldsymbol{\gamma} \leftarrow [0,0,0]$\;

\For{each minibatch $\mathcal{B} = \{x_i, y_i\}$}{

    $[\boldsymbol{\gamma}, \text{grad}] \leftarrow \text{PhaseSchedule}(t)$\;

    $[W_e, W_m, W_h] \leftarrow \text{softmax}(\boldsymbol{\gamma})$\;

    $e_i \leftarrow E(x_i)$\;
    \quad
    $s_i \leftarrow \max_{k} \cos(\theta_{i,k})$\;

    $\boldsymbol{\mu}_s, \boldsymbol{\sigma}_s 
    \leftarrow \text{UpdateStats}(\{s_i\}, 
    \boldsymbol{\mu}_s, \boldsymbol{\sigma}_s)$\;

    \For{each $i \in \mathcal{B}$}{
        $w_i \leftarrow
        \begin{cases}
            W_e & \text{if } s_i > \boldsymbol{\mu}_s + \boldsymbol{\sigma}_s \\
            W_h & \text{if } s_i < \boldsymbol{\mu}_s - \boldsymbol{\sigma}_s \\
            W_m & \text{otherwise}
        \end{cases}$\;
    }

    $\mathcal{L}_{\text{curr}} 
    \leftarrow \frac{1}{|\mathcal{B}|}
    \sum w_i \cdot L(e_i, y_i)$\;

    $\nabla_{\theta} \mathcal{L}_{\text{curry}}.\text{backward}()$\;
    \quad
    $\theta \leftarrow \text{Optimizer}(\theta)$\;
}
\end{algorithm}

All audio is resampled to 16~kHz and segmented into 3-second crops with random offset selection during training. To enhance model robustness, we apply augmentation dynamically during training: 
Adding Gaussian noise to the waveform by injecting white noise sampled from $\mathcal{N}(0, \sigma^2)$ with $\sigma \in [0.001, 0.015]$ uniformly drawn per utterance. The augmented waveforms
are converted into 160-dimensional log-Mel spectrograms using a 25~ms window and a 10~ms stride, and served as inputs to the speaker encoder. 
We train on three datasets spanning over 500K speaker identities. VoxCeleb1 and VoxCeleb2~\cite{nagrani2017voxceleb} together provide 7,000 speakers, comprising approximately 1.2 million clean utterances, and are loaded in full
each epoch. VoxBlink2~\cite{lin2024voxblink2} contributes 111K speakers sourced from
in-the-wild video, and CommonVoice~\cite{ardila2020common} provides over 340K speakers across English
and multilingual conditions originally collected for Automatic Speech Recognition (ASR). To manage epoch length while
preserving speaker diversity, VoxBlink2 and CommonVoice are resampled at 5 utterances
per speaker per epoch with a seed tied to the current epoch index, exposing different
utterances at each pass.

The model is trained end-to-end using AdamW~\cite{loshchilovdecoupled} with weight
decay $10^{-4}$ and a cosine learning rate schedule with linear warmup over 3 epochs.
We fully unfreeze the W2V-BERT~2.0 frontend and train it jointly with the MFA backend.
To protect the pre-trained SSL representations, we apply a differential learning rate:
the frontend is updated at $5 \times 10^{-6}$, while the randomly initialized backend,
classifier, and curriculum logits $\boldsymbol{\gamma}$ are updated at $5 \times
10^{-5}$, $5 \times 10^{-5}$, and $10^{-3}$, respectively. The Sub-center ArcFace
margin is progressively increased from $m=0.2$ to $m=0.35$ across training phases,
with feature scale $s=32$ and $K=3$ sub-centers throughout. All models are implemented in PyTorch~\cite{paszke2019pytorch} and trained on eight NVIDIA H100 GPUs with a per-GPU batch size of 128, yielding an effective batch size of 1024.
\begin{table}[t]
  \caption{EER (\%) and minDCF on VoxCeleb1-O and SITW. ($\downarrow$ better)}
  \label{tab:example}
  \centering
  \begin{tabular}{lccc}
    \toprule
    \textbf{Dataset} & \textbf{Loss} & \textbf{EER (\%)} & \textbf{minDCF} \\
    \midrule
    VoxCeleb1-O & Baseline & 2.87 & 0.45 \\
    VoxCeleb1-O & Curry & \textbf{0.38} & \textbf{0.04} \\
    \midrule
    SITW & Baseline & 4.00 & 0.27 \\
    SITW & Curry & \textbf{1.60} & \textbf{0.07} \\
    \bottomrule
  \end{tabular}
\end{table}

\section{Results and Analysis}
\begin{figure}[t]
\centering
\includegraphics[width=0.75\linewidth]{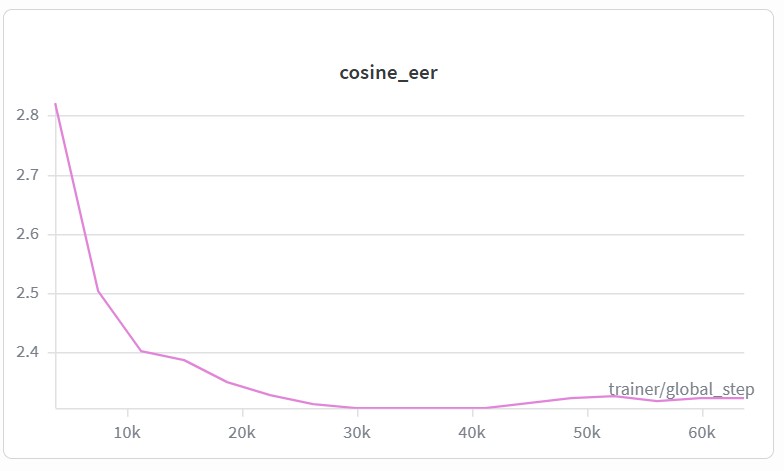}
\vspace{0.25em}
\includegraphics[width=0.75\linewidth]{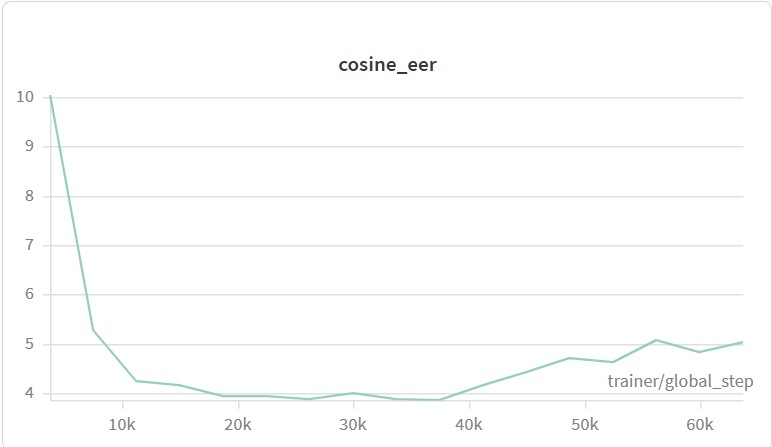}
\caption{Cosine EER (\%) evolution over training steps on a subset of the most difficult training instances.
\textit{Curry} (top) converges smoothly and stabilizes, while
the Sub-center ArcFace baseline (bottom) plateaus above 4\% and exhibits instability
in later epochs.}
\label{fig:eer_comparison}
\end{figure}

\begin{table}[t]
\caption{\label{tab:demographic_eer}EER (\%) across demographic subgroups in EARS dataset on speaker verficiation (↓ better)}
\centering
\begin{tabular}{llcccccc}
\hline
\textbf{Category} & \textbf{Subgroup} & \textbf{Baseline} & \textbf{Curry} \\
\hline
\multirow{2}{*}{Gender} 
& Female (59 spks) & 6.23 & \textbf{1.09} \\
& Male (43 spks)   & 9.95 & \textbf{1.84} \\
\hline
\multirow{11}{*}{Age} 
& F (18–25), 13 spks & 7.19 & \textbf{1.58} \\
& F (26–35), 13 spks & 6.34 & \textbf{1.15} \\
& F (36–45), 7 spks  & 4.41 & \textbf{0.40} \\
& F (46–55), 14 spks & 7.35 & \textbf{1.00} \\
& F (56–65), 10 spks & 7.78 & \textbf{0.88}  \\
& F (66–75), 2 spks  & 11.35 & \textbf{0.21}  \\
& M (18–25), 14 spks & 14.02 & \textbf{3.59}  \\
& M (26–35), 10 spks & 11.65 & \textbf{2.16}  \\
& M (36–45), 10 spks & 6.93 & \textbf{1.57}  \\
& M (46–55), 4 spks  & 8.24 & \textbf{2.57}  \\
& M (56–65), 5 spks  & 12.32 & \textbf{1.93}  \\
\hline
\end{tabular}
\end{table}
We benchmark our system on two standard speaker verification protocols. VoxCeleb1-O (Vox1-O)~\cite{nagrani2017voxceleb} comprises trials drawn from 40 speakers across celebrity interview recordings in diverse acoustic conditions. SITW~\cite{mclaren2016speakers} is a more challenging in-the-wild evaluation set consisting of approximately 1,000 verification pairs sourced from open-source media, covering a wide range of recording environments, microphone conditions, and vocal styles. As reported in Table~\ref{tab:example}, \textit{Curry} achieves \textbf{0.38\%} EER on Vox1-O and \textbf{1.60\%} on SITW, reducing the AAM-Softmax baseline by 86.8\% and 60.0\% relative, respectively. The minDCF improvements are equally consistent, dropping from 0.45 to 0.04 on Vox1-O and from 0.27 to 0.07 on SITW, confirming that the gains are not threshold-sensitive. Figure~\ref{fig:eer_comparison} further illustrates the convergence advantage of \textit{Curry}: while the baseline plateaus above 4\% EER and exhibits instability in later epochs, \textit{Curry} converges smoothly and stabilizes below 2.35\% within the first 30K steps on a subset of the most difficult training instances, consistent with our hypothesis that structured curriculum phasing suppresses gradient noise from degraded samples during the critical early stages of large-scale training.

To understand how speaker-discriminative information is distributed across demographic groups, following the \textsc{SVeritas}~\cite{baali2025sveritas} benchmark, we analyze model performance on the EARS~\cite{richter2024ears} dataset by splitting evaluation across gender and age subgroups, as reported in Table~\ref{tab:demographic_eer}.
\textit{Curry} consistently outperforms the baseline across all subgroups, reducing EER by up to 11$\times$ for female speakers aged 18--25 and maintaining strong performance across all age brackets. The baseline exhibits a marked gender disparity of 6.23\% vs. 9.95\% EER for female and male speakers respectively, a gap that \textit{Curry} substantially narrows to 1.09\% vs. 1.84\%, suggesting that difficulty-aware training implicitly mitigates the tendency of uniform-margin losses to overfit dominant acoustic conditions. Older male speakers (M~66--75) show the largest absolute improvement, from 11.35\% to 0.21\%, indicating that the progressive tier activation is particularly effective at handling the higher acoustic variability associated with underrepresented demographics.



\section{Conclusion}
In this paper we presented \textit{Curry} (CURriculum Ranking), an adaptive loss function that addresses the fundamental challenge of training speaker verification systems on
imperfect, large-scale data. By deriving per-sample confidence scores from sub-center
angular distances and dynamically partitioning each mini-batch into difficulty tiers
via running batch statistics, \textit{Curry} synchronizes gradient pressure with the
model's evolving representations; without auxiliary annotations or offline preprocessing. Its loss-agnostic design allows it to wrap any differentiable
per-sample objective, making it a general-purpose tool for large-scale speaker
training. Evaluated on VoxCeleb1-O and SITW, \textit{Curry} reduces EER by 86.8\%
and 60.0\% over the Sub-center ArcFace baseline, and demonstrates consistent robustness
gains across demographic subgroups on the \textsc{SVeritas} benchmark.
To our knowledge, this represents the largest-scale speaker verification system
trained to date, spanning 500K+ identities across diverse and imperfect data sources.

{\color{blue}
\section{Generative AI Use Disclosure}
Generative AI tools were used solely for polishing and editing the manuscript text,
as well as resizing tables and figures.
All technical content, experimental design, analysis, and conclusions are entirely
the work of the authors.
}

\bibliographystyle{IEEEtran}
\bibliography{mybib}

@inproceedings{deng2019arcface,
  title={Arcface: Additive angular margin loss for deep face recognition},
  author={Deng, Jiankang and Guo, Jia and Xue, Niannan and Zafeiriou, Stefanos},
  booktitle={Proceedings of the IEEE/CVF conference on computer vision and pattern recognition},
  pages={4690--4699},
  year={2019}
}

@article{lin2024voxblink2,
  title={Voxblink2: A 100k+ speaker recognition corpus and the open-set speaker-identification benchmark},
  author={Lin, Yuke and Cheng, Ming and Zhang, Fulin and Gao, Yingying and Zhang, Shilei and Li, Ming},
  journal={arXiv preprint arXiv:2407.11510},
  year={2024}
}

@article{nagrani2017voxceleb,
  title={VoxCeleb: A Large-Scale Speaker Identification Dataset},
  author={Nagrani, Arsha and Chung, Joon Son and Zisserman, Andrew},
  journal={Interspeech 2017},
  pages={2616},
  year={2017},
  publisher={ISCA}
}

@article{jung2022large,
  title={Large-scale learning of generalised representations for speaker recognition},
  author={Jung, Jee-weon and Heo, Hee-Soo and Lee, Bong-Jin and Lee, Jaesong and Shim, Hye-jin and Kwon, Youngki and Chung, Joon Son and Watanabe, Shinji},
  journal={arXiv preprint arXiv:2210.10985},
  year={2022}
}

@article{baali2025delulu,
  title={DELULU: Discriminative Embedding Learning Using Latent Units for Speaker-Aware Self-Trained Speech Foundational Model},
  author={Baali, Massa and Singh, Rita and Raj, Bhiksha},
  journal={arXiv preprint arXiv:2510.17662},
  year={2025}
}

@inproceedings{mclaren2016speakers,
  title={The speakers in the wild (SITW) speaker recognition database.},
  author={McLaren, Mitchell and Ferrer, Luciana and Castan, Diego and Lawson, Aaron},
  booktitle={Interspeech},
  pages={818--822},
  year={2016}
}

@inproceedings{baali2025sveritas,
  title={{SVeritas}: Benchmark for Robust Speaker Verification under Diverse Conditions},
  author={Baali, Massa and Bisht, Sarthak and Teixeira, Francisco and Shapovalenko, Kateryna and Singh, Rita and Raj, Bhiksha},
  booktitle={Findings of the Association for Computational Linguistics: EMNLP},
  year={2025}
}

@inproceedings{richter2024ears,
  title={{EARS}: An Anechoic Fullband Speech Dataset Benchmarked for Speech Enhancement and Dereverberation},
  author={Richter, Julius and Wu, Yi-Chiao and Krenn, Steven and Welker, Simon and Lay, Bunlong and Watanabe, Shinjii and Richard, Alexander and Gerkmann, Timo},
  booktitle={Interspeech},
  year={2024}
}

@inproceedings{baali2024pdaf,
  title={Pdaf: A phonetic debiasing attention framework for speaker verification},
  author={Baali, Massa and Aldoobi, Abdulhamid and Dhamyal, Hira and Singh, Rita and Raj, Bhiksha},
  booktitle={2024 IEEE Spoken Language Technology Workshop (SLT)},
  pages={1209--1216},
  year={2024},
  organization={IEEE}
}

@inproceedings{baali2025caarma,
  title={{CAARMA}: Class Augmentation with Adversarial Mixup Regularization},
  author={Baali, Massa and Li, Xiang and Chen, Hao and Hannan, Syed Abdul and Singh, Rita and Raj, Bhiksha},
  booktitle={Findings of the Association for Computational Linguistics: EMNLP},
  year={2025}
}

@inproceedings{yakovlev2024reshape,
  title={Reshape Dimensions Network for Speaker Recognition},
  author={Yakovlev, Ivan and Makarov, Rostislav and Balykin, Andrei and Malov, Pavel and Okhotnikov, Anton and Torgashov, Nikita},
  booktitle={Proc. Interspeech 2024},
  pages={3235--3239},
  year={2024}
}

@inproceedings{chung2021w2v,
  title={W2v-bert: Combining contrastive learning and masked language modeling for self-supervised speech pre-training},
  author={Chung, Yu-An and Zhang, Yu and Han, Wei and Chiu, Chung-Cheng and Qin, James and Pang, Ruoming and Wu, Yonghui},
  booktitle={IEEE Automatic Speech Recognition and Understanding Workshop},
  pages={244--250},
  year={2021},
  organization={IEEE}
}

@inproceedings{ardila2020common,
  title={Common voice: A massively-multilingual speech corpus},
  author={Ardila, Rosana and Branson, Megan and Davis, Kelly and Kohler, Michael and Meyer, Josh and Henretty, Michael and Morais, Reuben and Saunders, Lindsay and Tyers, Francis and Weber, Gregor},
  booktitle={Proceedings of the twelfth language resources and evaluation conference},
  pages={4218--4222},
  year={2020}
}

@article{paszke2019pytorch,
  title={Pytorch: An imperative style, high-performance deep learning library},
  author={Paszke, A. and Gross, S. and Massa, F. and Lerer, A. and Bradbury, J. and Chanan, G. and Killeen, T. and Lin, Z. and Gimelshein, N. and Antiga, L. and others},
  journal={Advances in neural information processing systems},
  volume={32},
  year={2019}
}

@inproceedings{loshchilovdecoupled,
  title={Decoupled Weight Decay Regularization},
  author={Loshchilov, Ilya and Hutter, Frank},
  booktitle={ICLR},
  year = {2019}
}

@inproceedings{deng2020sub,
  title={Sub-center arcface: Boosting face recognition by large-scale noisy web faces},
  author={Deng, Jiankang and Guo, Jia and Liu, Tongliang and Gong, Mingming and Zafeiriou, Stefanos},
  booktitle={European Conference on Computer Vision},
  pages={741--757},
  year={2020},
  organization={Springer}
}

@article{okabeattentive,
  title={Attentive Statistics Pooling for Deep Speaker Embedding},
  author={Okabe, Koji and Koshinaka, Takafumi and Shinoda, Koichi},
  booktitle={Interspeech},
  pages={2252–-2256},
  year={2018}
}

@article{chen2022wavlm,
  title={Wavlm: Large-scale self-supervised pre-training for full stack speech processing},
  author={Chen, Sanyuan and Wang, Chengyi and Chen, Zhengyang and Wu, Yu and Liu, Shujie and Chen, Zhuo and Li, Jinyu and Kanda, Naoyuki and Yoshioka, Takuya and Xiao, Xiong and others},
  journal={IEEE Journal of Selected Topics in Signal Processing},
  volume={16},
  number={6},
  pages={1505--1518},
  year={2022},
  publisher={IEEE}
}

@article{li2025enhancing,
  title={Enhancing Speaker Verification with w2v-BERT 2.0 and Knowledge Distillation guided Structured Pruning},
  author={Li, Ze and Cheng, Ming and Li, Ming},
  journal={arXiv preprint arXiv:2510.04213},
  year={2025}
}

@inproceedings{vera2015speaker,
  title={Speaker identification system based in verification techniques with Bayesian discrimination},
  author={Vera, M. and Pelle, P. and Estienne, C. and Ferrer, L.},
  booktitle={2015 XVI Workshop on Information Processing and Control (RPIC)},
  pages={1--6},
  year={2015},
  organization={IEEE}
}

@inproceedings{snyder2018x,
  title={X-vectors: Robust dnn embeddings for speaker recognition},
  author={Snyder, D. and Garcia-Romero, D. and Sell, G. and Povey, D. and Khudanpur, S.},
  booktitle={ICASSP},
  pages={5329--5333},
  year={2018},
  organization={IEEE}
}

@inproceedings{doddipatla2017speaker,
  title={Speaker Adaptation in DNN-Based Speech Synthesis Using d-Vectors.},
  author={Doddipatla, R. and Braunschweiler, N. and Maia, R.},
  booktitle={Interspeech},
  pages={3404--3408},
  year={2017}
}

@article{singh2025human,
  title={Human voice is unique},
  author={Singh, Rita and Raj, Bhiksha},
  journal={arXiv preprint arXiv:2506.18182},
  year={2025}
}

@article{ahmed2026quantifying,
  title={Quantifying the Relationship Between Speech Quality Metrics and Biometric Speaker Recognition Performance Under Acoustic Degradation},
  author={Ahmed, Ajan and Imtiaz, Masudul H},
  journal={Signals},
  volume={7},
  number={1},
  pages={7},
  year={2026},
  publisher={MDPI}
}

@article{hintz2024commonbench,
  title={CommonBench: A larger scale speaker verification benchmark},
  author={Hintz, Jan and Siegert, Ingo},
  journal={Proc. SPSC},
  volume={2024},
  pages={17--20},
  year={2024}
}

@article{farhadipour2026tidyvoice,
  title={TidyVoice: A Curated Multilingual Dataset for Speaker Verification Derived from Common Voice},
  author={Farhadipour, Aref and Marquenie, Jan and Madikeri, Srikanth and Chodroff, Eleanor},
  journal={arXiv preprint arXiv:2601.16358},
  year={2026}
}

@inproceedings{bengio2009curriculum,
  title={Curriculum learning},
  author={Bengio, Yoshua and Louradour, J{\'e}r{\^o}me and Collobert, Ronan and Weston, Jason},
  booktitle={Proceedings of the 26th annual international conference on machine learning},
  pages={41--48},
  year={2009}
}

@article{ranjan2017curriculum,
  title={Curriculum learning based approaches for noise robust speaker recognition},
  author={Ranjan, Shivesh and Hansen, John HL},
  journal={IEEE/ACM Transactions on Audio, Speech, and Language Processing},
  volume={26},
  number={1},
  pages={197--210},
  year={2017},
  publisher={IEEE}
}

@article{heo2022curriculum,
  title={Curriculum learning for self-supervised speaker verification},
  author={Heo, Hee-Soo and Jung, Jee-weon and Kang, Jingu and Kwon, Youngki and Kim, You Jin and Lee, Bong-Jin and Chung, Joon Son},
  journal={arXiv preprint arXiv:2203.14525},
  year={2022}
}

@article{bai2022end,
  title={End-to-end speaker verification via curriculum bipartite ranking weighted binary cross-entropy},
  author={Bai, Zhongxin and Wang, Jianyu and Zhang, Xiao-Lei and Chen, Jingdong},
  journal={IEEE/ACM Transactions on Audio, Speech, and Language Processing},
  volume={30},
  pages={1330--1344},
  year={2022},
  publisher={IEEE}
}



\end{document}